\documentclass[manuscript]{acmart}
\usepackage{xspace}
\usepackage{tabularx}
\usepackage{amsmath}
\usepackage{soul}
\usepackage{caption}
\usepackage{subcaption}
\usepackage{enumitem}

\AtBeginDocument{%
  \providecommand\BibTeX{{%
    \normalfont B\kern-0.5em{\scshape i\kern-0.25em b}\kern-0.8em\TeX}}}

\copyrightyear{2021} 
\acmYear{2021} 
\setcopyright{acmlicensed}\acmConference[CHI '21 Extended Abstracts]{CHI Conference on Human Factors in Computing Systems Extended Abstracts}{May 8--13, 2021}{Yokohama, Japan}
\acmBooktitle{CHI Conference on Human Factors in Computing Systems Extended Abstracts (CHI '21 Extended Abstracts), May 8--13, 2021, Yokohama, Japan}
\acmPrice{15.00}
\acmDOI{10.1145/3411763.3451618}
\acmISBN{978-1-4503-8095-9/21/05}

\begin{document}

\newcolumntype{L}[1]{>{\raggedright\let\newline\\\arraybackslash\hspace{0pt}}p{#1}}

\newcommand{\rulesep}{\unskip\ \textcolor{gray}{\vrule}\ }

\title{A bibliometric analysis of citation diversity in accessibility and HCI research}

\author{Lucy Lu Wang}
\affiliation{%
  \institution{Allen Institute for AI}
  \streetaddress{}
  \city{Seattle}
  \country{USA}}
\email{lucyw@allenai.org}

\author{Kelly Mack}
\affiliation{%
  \institution{University of Washington}
  \streetaddress{}
  \city{Seattle}
  \country{USA}}
\email{kmack3@uw.edu}

\author{Emma McDonnell}
\affiliation{%
  \institution{University of Washington}
  \streetaddress{}
  \city{Seattle}
  \country{USA}}
\email{ejm249@uw.edu}

\author{Dhruv Jain}
\affiliation{%
  \institution{University of Washington}
  \streetaddress{}
  \city{Seattle}
  \country{USA}}
\email{djain@uw.edu}

\author{Leah Findlater}
\affiliation{%
  \institution{University of Washington}
  \streetaddress{}
  \city{Seattle}
  \country{USA}}
\email{leahf@uw.edu}

\author{Jon E. Froehlich}
\affiliation{%
  \institution{University of Washington}
  \streetaddress{}
  \city{Seattle}
  \country{USA}}
\email{jonf@cs.uw.edu}

\renewcommand{\shortauthors}{Wang LL et al. 2021}

\begin{abstract}
  Accessibility research sits at the junction of several disciplines, drawing influence from HCI, disability studies, psychology, education, and more. To characterize the influences and extensions of accessibility research, we undertake a study of citation trends for accessibility and related HCI communities. We assess the diversity of venues and fields of study represented among the referenced and citing papers of 836 accessibility research papers from ASSETS and CHI, finding that though publications in computer science dominate these citation relationships, the relative proportion of citations from papers on psychology and medicine has grown over time. Though ASSETS is a more niche venue than CHI in terms of citational diversity, both conferences display standard levels of diversity among their incoming and outgoing citations when analyzed in the context of 53K papers from 13 accessibility and HCI conference venues. 
\end{abstract}






\maketitle

\section{Introduction}
\label{sec:introduction}

Accessibility is an increasingly prominent area of research, one which identifies, assesses, and innovates to improve upon the accessibility challenges in computing technologies. The field of accessibility is closely tied to and influenced by human-computer interaction (HCI), disability studies, education, and more, and ideas are frequently borrowed and shared among these disciplines. In this work, we examine the relationship between accessibility research and these connected fields using bibliometric and citation analysis methods. By directing an analytical lens back on ourselves, the community can better reflect upon the impacts of its work and identify ways to increase interdisciplinary collaboration in a meaningful way. The goal of this work is to answer the following questions:

\begin{itemize}[noitemsep,topsep=1pt]
    \item What are the citation patterns of accessibility research published at CHI and ASSETS?
    \item How does accessibility research relate to other computing fields and to fields outside of computing, e.g., what are the incoming and outgoing citation patterns within and beyond accessible computing?
    \item What are the trends over time and how are they evolving?
    \item How do these patterns and trends compare to other communities in HCI?
\end{itemize}

To address these questions, we perform exploratory analysis to examine the citation counts and citation diversity in the accessibility research community.\footnote{Code and data for this paper are available at \href{https://github.com/makeabilitylab/accessibility-bibliometric-analysis}{https://github.com/makeabilitylab/accessibility-bibliometric-analysis}} Citation diversity provides an indicator of relationships between different fields. To characterize citation diversity, we assess the most common publication venues and fields of study among the references and citations (where \textit{references} refer to outbound citations and \textit{citations} to incoming citations) of 836 accessibility publications identified by \citet{mack_whatdw_chi_2021}, along with how trends in citations to HCI and other fields have changed over time. For comparison, we analyze the citational diversity of 53K papers from 13 accessibility and HCI conferences using the Leinster–Cobbold diversity index (LCDI) \citep{Leinster2012MeasuringDT} to contextualize the diversity of the field of accessibility within the greater ecosystem of HCI.

We show that accessibility papers published in CHI versus ASSETS demonstrate similar citation features. Median citations received by CHI and ASSETS papers are similar: 23 and 24 respectively. Venues and fields of study represented among the references and citations of these papers are also similar, though the diversity of fields of study is higher among references than citations for both conferences. The primary non-HCI venues citing work in accessibility are those in the related areas of physical medicine and rehabilitation engineering. We find that though accessibility papers published in CHI or ASSETS do not appear to have substantially different citational outcomes, the LCDI diversity measure shows a clear distinction between the overall conferences (computed for all papers versus just accessibility papers). Accessibility researchers will find complementary benefits to publishing in both venues, through interaction with the more thematic community at ASSETS and the broader, more interdisciplinary community at CHI.

\section{Related work}
\label{sec:related_work}
Bibliometric analysis has been used broadly to study patterns in authorship, citation, and collaboration in scientific publishing \citep{Holman2018TheGG, Ribeiro2017GrowthPO, Kas2011TrendsIS, Newman2001TheSO}. Many studies have investigated the role of different paper features (open access \citep{Gargouri2010SelfSelectedOM}, preprint availability \citep{Feldman2018CitationCA}, social media amplification \citep{Klar2020UsingSM, Haustein2015CharacterizingSM}, \textit{etc.}) on citation count. Though citations are correlated with some perceptions of a paper's success, they are an imperfect measure of importance and influence. Rather than relying on raw citation count alone, we assess the interdisciplinarity of accessibility research. Prior work has measured scientific interdisciplinarity based on the diversity of a paper's outgoing references, exploring diversity indices like LCDI \citep{Zhang2016DiversityOR,
Mugabushaka2016BibliometricIO} or Rao-Stirling \citep{LEYDESDORFF201187}, and network features derived from the collaboration and authorship graph \citep{LEYDESDORFF201187, Karlovcec2014InterdisciplinarityOS}. \citet{Zhang2016DiversityOR} compute aggregate LCDI as an indicator of a journal's interdisciplinarity, which we adopt to assess venue interdisciplinarity. In this work, we assess the diversity of venues and fields among the referenced and citing papers of accessibility research, and compute aggregate LCDI for several HCI publication venues, using these metrics to characterize the interdisciplinarity of CHI and ASSETS in the context of other HCI venues.

Bibliometric methods have been used to survey papers in computing, with several studies conducted on HCI research to identify emerging trends \citep{Koumaditis2017HumanCI} and to study patterns in paper authorship or citations \citep{guo_bibliometricao_2020, Marshall2017ThrowawayCO, Coursaris2012AMR, Dillon1995MappingTD}. In several cases, authors have applied these methods to better understand the impacts of papers published in specific HCI publication venues, like \textit{IJHCS} and \textit{CHI} \citep{Mannocci2019TheEO}, or \textit{Human Factors} \citep{Lee2005BibliometricAO}. By providing a top-down overview of the state of a field, these bibliometric reviews can provide jumping-off points for new ideas, especially for researchers first entering a field. In our case, we focus on citation analysis as one way of assessing the interdisciplinarity of the accessible computing community, where as far as we know, such analyses have not been conducted before.

\begin{table*}[t!]
    \centering
    \small
    \begin{tabular}{llll}
        \toprule
        Venue & Active years & Entries in DBLP & Full name of conference \\
        \midrule
        ASSETS & 1994-2020 & 1355 & ACM SIGACCESS Conference on Computers and Accessibility \\ 
        CHI & 1981-2020 & 16446 & ACM Conference on Human Factors in Computing Systems \\
        HCI & 1987-2020 & 17521 & International Conference on Human-Computer Interaction \\
        UbiComp & 2001-2020 & 3267 & ACM Conference on Ubiquitous Computing \\
        CSCW & 1986-2020 & 2537 & ACM Conference on Computer Supported Cooperative Work\\
        IUI & 1993-2020 & 2028 & ACM Conference on Intelligent User Interfaces \\
        UIST & 1988-2020 & 1927 & ACM Symposium on User Interface Software and Technology \\
        ICCHP & 1994-2020 & 1748 & International Conference on Computers Helping People with Special Needs \\
        DIS & 1995-2020 & 1476 & ACM Conference on Designing Interactive Systems \\
        OZCHI & 2005-2019 & 1264 & Australian Conference on Human-Computer Interaction \\
        TEI & 2007-2020 & 1210 & ACM Conference on Tangible and Embedded Interaction \\
        IDC & 2003-2020 & 1193 & ACM Conference on Interaction Design and Children \\
        NordiCHI & 2002-2020 & 995 & Nordic Conference on Human-Computer Interaction \\
        \bottomrule
    \end{tabular}
    \caption{Venues for comparison. Note that several conferences occurred biennially or irregularly within some year ranges.}
    \label{tab:all_venues}
\end{table*}

\section{Data \& Methods}
\label{sec:methods}

We leverage the open dataset of accessibility papers released by Mack et al.~\citep{mack_whatdw_chi_2021}. This dataset includes 836 accessibility papers from the CHI and ASSETS conferences (260 CHI and 576 ASSETS papers) spanning 1994--2019\footnote{Dataset available at \href{https://github.com/makeabilitylab/accessibility-literature-survey}{https://github.com/makeabilitylab/accessibility-literature-survey}}; all papers were manually curated by the authors. We refer to these 836 papers as our \textbf{\textit{core}} set. We call documents referenced by these papers (outbound citations) as \textit{references} and documents citing these papers (inbound citations) as \textit{citations}.

To better understand the relationship between accessibility / accessible computing and other fields of study, we assess the publication venues of the references and citations of the core set, using venue as a coarse proxy for scientific community. We also analyze each document's field of study as classified by the Microsoft Academic search engine \citep{msr:mag1,Shen2018AWS}, which offers better insight into the distribution of topics discussed in these documents. For context, we construct a \textbf{\textit{comparative}} dataset of 53K publications from 13 selected conferences in accessibility and HCI (including ASSETS and CHI) along with their references and citations. Field of study diversity analysis is performed on the references and citations of this comparative set to help guide interpretation of citation diversity amongst the core accessibility set. Table \ref{tab:all_venues} provides a list of selected comparison venues, along with statistics on document counts and publication history. 

\subsection{Dataset construction}
\label{sec:dataset}

Metadata for all papers are derived from DBLP \citep{Ley2009DBLPS}, Semantic Scholar \citep{Ammar2018ConstructionOT}, and Microsoft Academic \citep{msr:mag1}. Since no database of computer science publications is complete or even particularly comprehensive \citep{Cavacini2014WhatIT}, we select DBLP as the primary source of paper metadata because of its emphasis on manual curation and quality \footnote{See \href{https://dblp.org/faq/5210119.html}{https://dblp.org/faq/5210119.html} for inclusion criteria and \href{https://dblp.org/faq/13500484.html}{https://dblp.org/faq/13500484.html} for DBLP's data curation workflow.} as well as its high coverage of HCI venues. We derive digital object identifiers (DOIs), publication year, and normalized publication venues from DBLP \citep{Ley2009DBLPS}. We derive citations and references for each paper in the core set using the Semantic Scholar API \citep{Ammar2018ConstructionOT}. The 836 core papers reference 21464 documents (14184 unique) and are cited by 30355 documents (17208 unique). Unsurprisingly, 750 (89.7\%) of the 836 papers in the core set are cited by another paper in the core set.

We derive metadata for the references and citations by linking them to DBLP or Semantic Scholar. Together, 22830 (77.6\%) of the 29410 unique referenced or cited documents have DOIs.\footnote{DOIs are provided by most large academic publishers, and are the most widely used identifiers for scholarly publications. However, not all publications receive DOIs, e.g. some conferences and workshops do not acquire them, some books may only have ISBNs, etc. A coverage of 78\% is fairly standard.} 
We use DOIs to link 11035 (51.4\%) references and 17203 (56.7\%) citations to DBLP, from which we derive normalized venue metadata. An additional 4464 references and 3783 citations are linked to Semantic Scholar; the venue data in Semantic Scholar is not normalized---i.e. the venue is a string value that must be mapped to a normalized venue, e.g., ``CHI '19'' and ``The 2019 ACM CHI Conference on Human Factors in Computing Systems'' must both be mapped to \textit{CHI}. We heuristically and manually map these venue strings to normalized venues. We are unable to find venue information for 2658 unique references and 4760 unique citations. Most (71.5\%) of these venue-less documents lack DOIs, making them challenging to identify or link. Of those with DOIs, we investigate a sample to better understand what they are and how their lack of venue information could impact our analysis. An assessment of the 100 most commonly occurring DOIs within this set reveals that most of these (73 of 100) resolve to books, book chapters, reports, or other document types without associated venues. Of the documents that have a publication venue unknown to DBLP or Semantic Scholar (23 of 100), all are from less well-known venues, and none are associated with the venues selected for our analysis. Therefore, we anticipate minimal bias to our results due to missing venue data. Details on this error analysis and additional commentary are available in Appendix~\ref{app:error_analysis}. 

Finally, we map all papers to the Microsoft Academic Graph (MAG) to derive their fields of study \citep{Shen2018AWS, msr:mag1}. The MAG fields are organized into a six-level hierarchy, and we retain and analyze all fields in the upper two levels (L0-L1). L0 is the highest level, and includes 19 fields such as Medicine, Psychology, and Computer Science. L1 fields are more granular, including things like human computer interaction, computer vision, developmental psychology, physical therapy, etc. Though the hierarchy continues into L2 and beyond, the fields quickly become too specific, which is why we elect to perform analysis over only the top two levels. Each paper can be associated with multiple fields of study at each level. Each field of study may have multiple parents, though we default to selecting a primary parent when displaying the L0 information associated with any particular L1 field. Of the 21464 references and 30355 citations of the core set, we identify field of study information for 19252 (89.7\%) references and 26997 (88.9\%) citations.

To provide context for interpretation, we select a set of 13 conferences that publish accessibility and HCI research (including ASSETS and CHI) for comparison. These publication venues (Table \ref{tab:all_venues}) are selected based on proximity and prestige to accessibility and HCI. They include general HCI venues like CHI and HCI, sub-discipline specific venues like TEI and UIST, as well as regional conferences like OZCHI and NordiCHI that are similar in size to ASSETS. Note that the number of entries for ASSETS in Table \ref{tab:all_venues} is much higher than the paper count in the core dataset, which includes all \textit{full-length} accessibility papers at both ASSETS and CHI and no extended abstracts. In contrast, Table \ref{tab:all_venues} is derived from DBLP and includes full length papers along with other types of accepted submissions such as posters, late-breaking work, and/or demos. Given this distinction, none of these venues are directly comparable to the core set, and are rather used to provide context for the expected reference and citation diversities in similar venues. Similar to the core set, references and citations for papers in these comparative venues are derived from Semantic Scholar, and venue and field of study information from DBLP, Semantic Scholar, and MAG as previously described.

\subsection{Analyses}
\label{sec:analyses}

We examine citation patterns in accessibility research, references and citations to/from other fields, and temporal trends. \\ [-4mm]

\noindent \textit{Analysis of venues}: We aggregate all references and citations for the core dataset. Identifying the venues of these papers, we then determine the top venues referenced by and citing these accessibility papers. Venue is one of two proxy measures we use to distinguish between research communities. \\ [-4mm]

\noindent \textit{Analysis of fields of study}: We analyze the MAG L1 fields of study of each paper referenced by or citing a paper in the core set. We also analyze temporal trends to determine whether the proportional representation of certain fields is increasing, decreasing, or remaining stable over time. Though we refer to these as fields based on MAG's nomenclature, each field more closely resembles a topic. Therefore, though a paper may be published in a computer science venue like CHI, it may be about a combination of topics, including ones in computer science like human-computer interaction or computer security, but also outside of computer science like epistemology or ethics. \\ [-4mm]

\noindent  \textit{Comparative analysis}: Finally, we perform field of study analysis across all 13 comparative venues. To compare the diversity among referenced and citing papers of these venues, we compute a diversity index over their MAG fields of study. The LCDI is computed over the L1 fields for the references and citations of each paper, and is defined as:
\begin{equation}
    \displaystyle LCDI = \left(\displaystyle \sum\limits_{i,j=1}^N s_{ij} p_i p_j\right)^{-1}
\end{equation} 

\noindent where $s_{ij}$ gives the similarity between two fields of study, $p_i$ is the proportion of references in field $i$ out of $N$ total fields, and $j$ the field of the paper of interest. We derive the similarity $s_{ij}$ between fields using the hierarchy defined by MAG as $\frac{1}{2^n}$ where $n$ is the number of levels of hierarchy that must be traversed to find a common parent. The larger the LCDI, the more diverse the fields of study are among the reference or citation pool for that paper. In the case where all referenced papers are in the same field of study as the paper of interest, the LCDI equals 1. For each comparative venue, we compute the LCDI of all references and citations, and compare the distributions of these diversity scores. 

\section{Results}
\label{results}

\begin{figure*}[t!]
  \centering
  \includegraphics[width=0.45\linewidth]{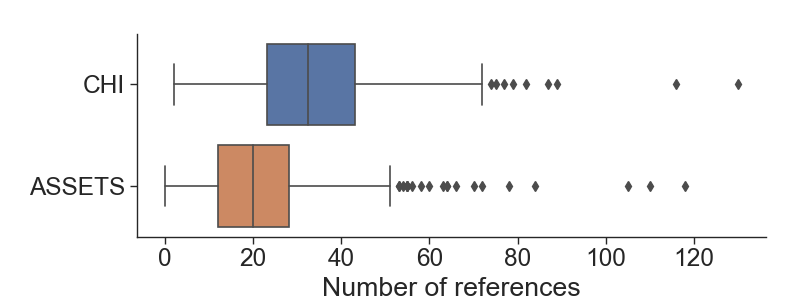}\includegraphics[width=0.45\linewidth]{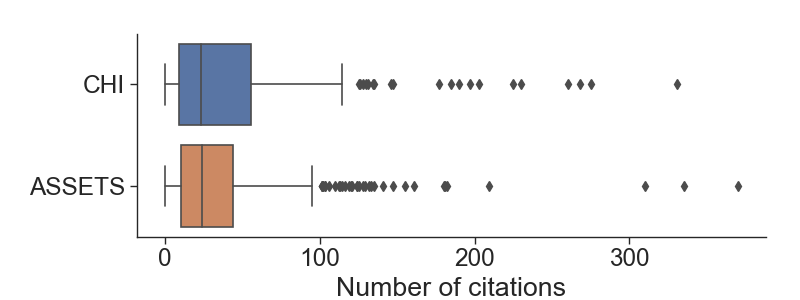}
  \caption{Distribution of reference (outbound citations) and citation (incoming citations) counts for accessibility papers in the core set split by ASSETS (N=576) and CHI (N=260). CHI accessibility papers have a median of 32.5 references (mean=34.5; SD=18.3) and ASSETS 20.0 references (mean=21.7; SD=14.7). For citations, CHI accessibility papers have a median of 23.0 citations (mean=41.9; SD=52.1) and ASSETS 24.0 citations (mean=33.8; SD=38.9).}
  \label{fig:summary}
  \Description{Two box plots show the distribution of the number of references per paper and citations per paper, respectively, for the core set, split by CHI and ASSETS. CHI references (median = 32.5; mean = 34.5; SD = 18.3). ASSETS references (median = 20.0; mean = 21.7; SD = 14.7). CHI citations (median = 23.0; mean = 41.9; SD= 52.1). ASSETS citations (median = 24.0; mean = 33.8; SD = 38.9).}
\end{figure*}

The 836 papers in the core accessibility set reference a median of 23 papers (mean = 25.7; SD = 17.0) and are cited by a median of 24 papers (mean = 36.3; SD = 43.6). Figure \ref{fig:summary} shows the distribution of reference and citation counts per paper in the core dataset split by venue. The average number of references is much higher for CHI (mean = 34.5; SD = 18.3) than ASSETS (mean = 21.7; SD = 14.7). Though the average citation count is also higher for CHI (mean = 41.9; SD = 52.1) than ASSETS (mean = 33.8; SD = 38.9), the median is similar for both venues, 23 for CHI and 24 for ASSETS.

\begin{figure*}[t!]
  \centering
  \includegraphics[width=0.43\linewidth]{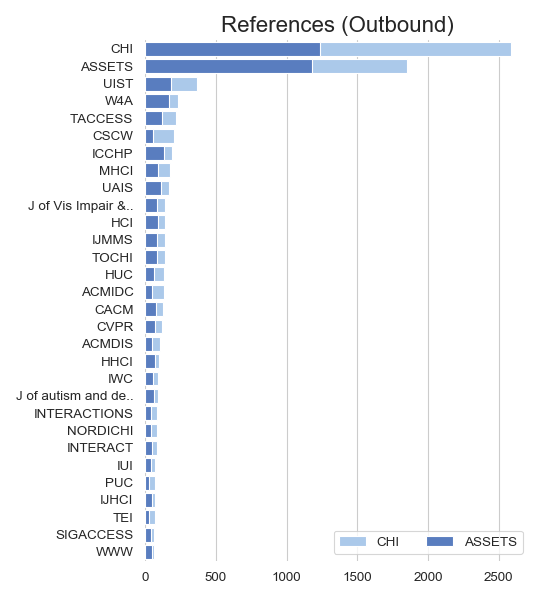}\includegraphics[width=0.43\linewidth]{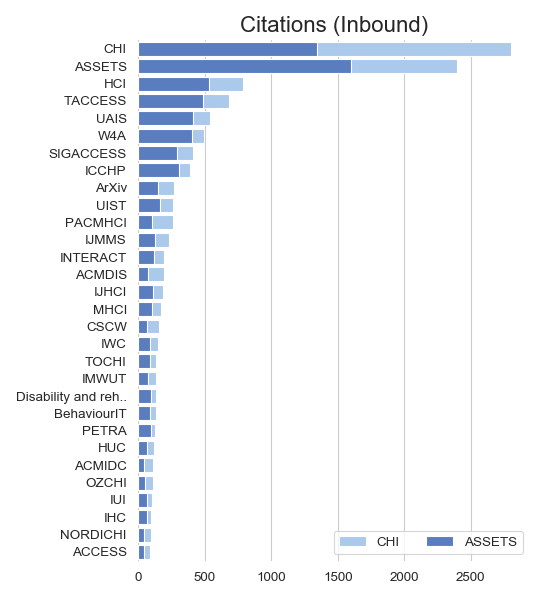}
  \caption{Top venues of papers referenced by (\textit{left}) and citing (\textit{right}) accessibility papers in the core set. References and citations are both dominated by papers from CHI and ASSETS, though a relatively larger proportion of citations arrive from other publication venues.}
  \label{fig:top_a11y_venues}
  \Description{Two bar plots show the top occurring venues among the references and citations, respectively, of our core set, split by CHI and ASSETS. Both references and citations are dominates by works published at CHI and ASSETS. In decreasing order, references are from CHI, ASSETS, UIST, W4A, TACCESS, CSCW, ICCHP, MHCI, UAIS, Journal of Visual Impair \& Blindness, HCI, IJMMS, TOCHI, etc. In decreasing order, citations are from CHI, ASSETS, HCI, TACCESS, UAIS, W4A, SIGACCESS, ICCHP, ArXiv, UIST, PACMHCI, IJMMS, INTERACT, etc. ASSETS accessibility papers are equally likely to reference CHI and ASSETS papers, but CHI accessibility papers are about twice as likely to reference CHI papers than ASSETS papers. CHI accessibility papers receive more citations from CHI than ASSETS, and ASSETS more from ASSETS than CHI.}
\end{figure*}

Figure \ref{fig:top_a11y_venues} shows the top venues represented among referenced and citing papers. CHI and ASSETS papers make up a substantial portion of references and citations, and are especially well-represented among references. For references, ASSETS papers cite a similar number of papers in CHI and ASSETS, though CHI papers in our core set are around twice as likely to cite CHI papers as ASSETS papers ($\chi^2 = 108.1$, $p < 0.001$). Other reference behaviors are similar between the two subsets, though CHI papers are more likely to cite papers published in CSCW and IDC. Citations, on the other hand, are more likely to come from papers in the same venue, i.e., citations to ASSETS papers are more likely to come from ASSETS papers, and citations to CHI papers from CHI papers ($\chi^2 = 182.4$, $p < 0.001$) (see contingency tables in Appendix~\ref{app:contingency_tables}). The most commonly occurring non-computer science venues among references are disability-related journals like the \textit{Journal of Visual Impairment \& Blindness} (141 references) and \textit{Journal of Autism and Developmental Disorders} (89 references), and among citations, \textit{Disability and Rehabilitation: Assistive Technology} (133 citations) and \textit{IEEE Trans. Neural Systems \& Rehab. Engineering} (60 citations).

\begin{figure*}[t!]
  \centering
  \includegraphics[width=0.49\linewidth]{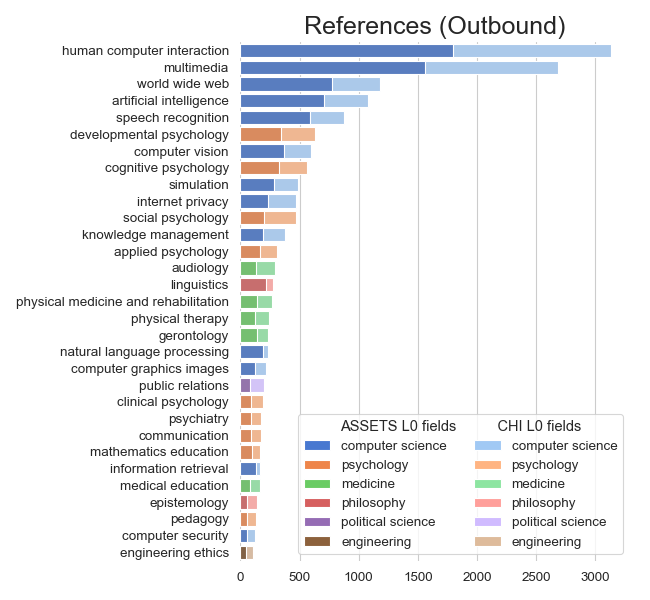}\includegraphics[width=0.49\linewidth]{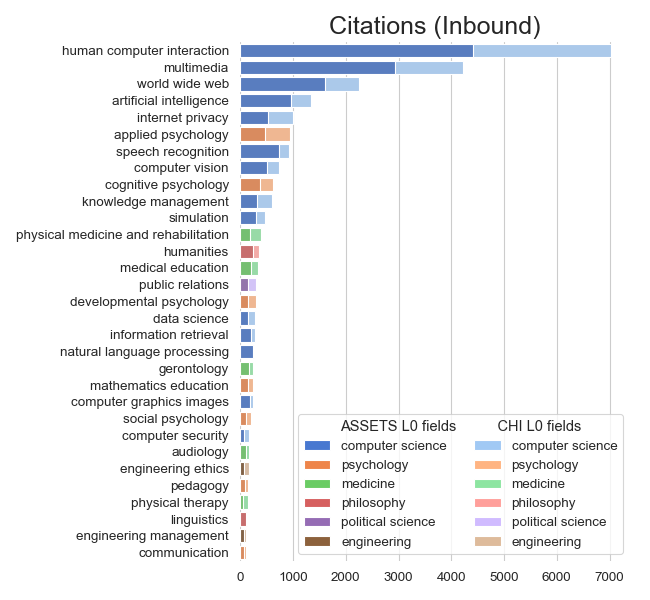}
  \caption{Top L1 fields of study of papers referenced by (\textit{left}) and citing (\textit{right}) accessibility papers in the core set. The primary color of each bar is determined by the L0 parent of that field, e.g., subfields of computer science are blue and subfields of psychology are orange. References show a greater diversity of fields of study (especially those outside of computer science) compared to citations.}
  \label{fig:top_a11y_fos}
  \Description{Two bar plots show the top occurring L1 fields of study among the references and citations, respectively, of our core set, split by CHI and ASSETS. Both references and citations are dominates by works in the field of human-computer interaction. In decreasing order, references are in the fields of human computer interaction, multimedia, world wide wide, artificial intelligence, speech recognition, developmental psychology, computer vision, cognitive psychology, simulation, internet privacy, social psychology, etc. In decreasing order, citations are in the fields of human computer interaction, multimedia, world wide wide, artificial intelligence, internet privacy, applied psychology, speech recognition, computer vision, cognitive psychology, knowledge management, simulation, etc. ASSETS and CHI do not differ much on referenced and citing fields, though the representation of fields from outside computer science is much higher among references than citations.}
\end{figure*}

\begin{figure*}[t!]
  \centering
    \includegraphics[width=0.32\linewidth]{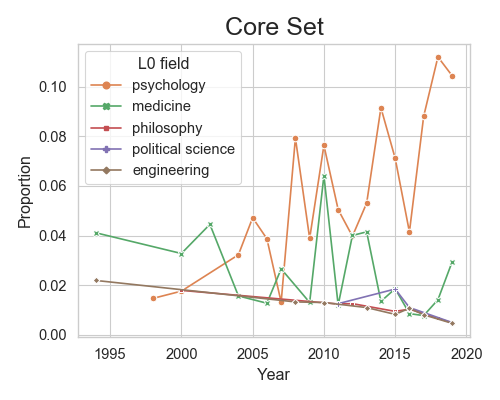}
    \rulesep
    \includegraphics[width=0.32\linewidth]{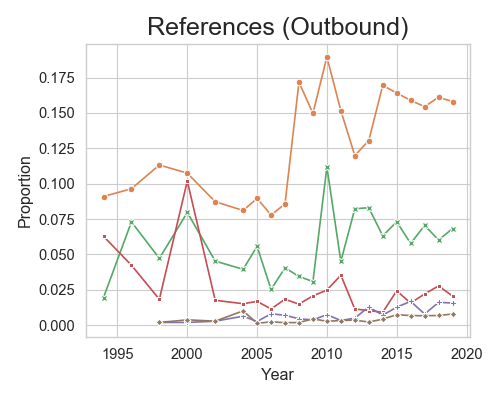}
    \includegraphics[width=0.32\linewidth]{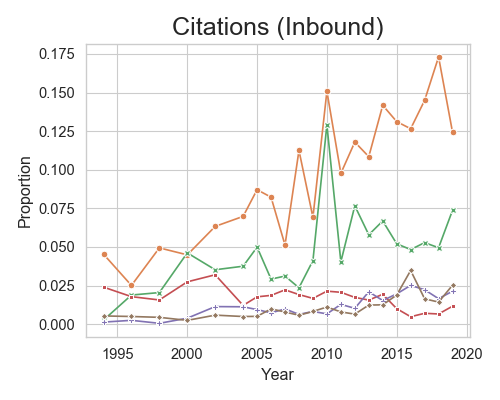}
  \caption{Proportions of non-CS fields represented among papers in the core set over time (\textit{left}), and among the papers referenced by (\textit{center}) and citing (\textit{right}) papers in the core set. The proportions of papers on the topic of psychology, and medicine to a lesser degree, have increased over time, especially among citations. Only years with greater than 5 papers are shown.}
  \label{fig:fos_over_time}
  \Description{A first line plot shows trends in non-computer science fields of study over time for the core set. Psychology is increasing the most from around 2\% in 1998 to 10\% in 2019. Medicine is flat or decreasing from around 4\% in 1995 to 3\% in 2019. Philosophy, political science, and engineering all have low representation, decreasing from 2\% to around 1\%. Two other line plots show trends in non-computer science fields of study over time for the references and citations of the core set, respectively. The representation of psychology has increased over time, especially for citations. The proportion of papers on medicine has also increased over time, to a lesser degree.}
\end{figure*}

\begin{figure*}[t!]
  \centering
  \includegraphics[width=0.36\linewidth]{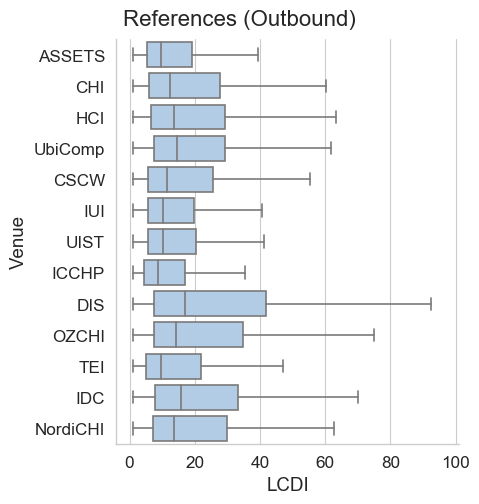}\hspace{6mm}\includegraphics[width=0.36\linewidth]{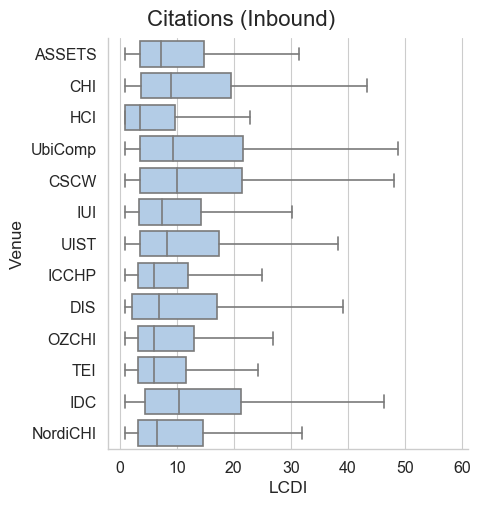}
  \caption{Distribution of the LCDI \citep{Leinster2012MeasuringDT} computed over the L1 fields of study of referenced papers (\textit{left}) and citing papers (\textit{right}) for each of 13 venues in the comparative dataset. Higher LCDI indicates higher diversity among the fields of the references or citations. LCDI is similar across venues but is lower for those focused on a particular subdomain (e.g. ASSETS, IUI, ICCHP, TEI) or regional conferences (OZCHI, NordiCHI) and higher for more general HCI conferences (CHI, UbiComp).}
  \label{fig:lcdi_box}
  \Description{Two sets of box plots show the distribution of LCDI for the references and citations, respectively, of all publications in 13 comparative accessibility and HCI-related venues. CHI as a whole (beyond just accessibility papers) demonstrates greater diversity (LCDI) among both references and citations compared to ASSETS.}
\end{figure*}

Figure~\ref{fig:top_a11y_fos} shows the distribution of fields among referenced and citing papers for the core set, split by CHI and ASSETS. Papers referenced by accessibility papers tend to be about a more diverse set of topics (relative counts of non-HCI papers are higher among references than citations). Though these papers are dominated by computer science subfields, many subfields of psychology and medicine are represented. Citations are more niche, originating predominantly from topics in computer science. We aggregate the non-CS subfields among the top 30 fields shown in Figure \ref{fig:top_a11y_fos} into their L0 parent fields, and present temporal trends for these aggregate L0 fields in Figure~\ref{fig:fos_over_time}. Over time, papers on the topic of psychology appear more frequently among both references and citations, and to a lesser degree, we see an increasing proportion of citations from papers on the topic of medicine. These increases can be partially attributed to the growing diversity of fields represented by papers in the core set, e.g., in recent years, more papers in the core set are being classified into subfields of psychology. We show this change in Figure~\ref{fig:fos_over_time} (left), which plots the distribution of non-CS L0 fields associated with L1 fields seen among the papers of the core set. A reference or citation to an accessibility paper that is classified into a subfield of psychology will artificially inflate the representation of psychology among the referenced and citing papers. However, this increase is not seen for medicine, and the increase in incoming citations to accessibility papers from papers on subtopics of medicine may derive from other sources. 

Figure~\ref{fig:lcdi_box} shows the distribution of the LCDI for each venue. As expected, more general HCI venues like CHI and UbiComp have higher diversity among references and citations than subdomain-focused venues like ASSETS, ICCHP, or TEI. Among these venues, papers in CHI, UbiComp, CSCW, and IDC influence the most diverse set of papers. These LCDI values show that the citation diversity for these HCI venues is generally lower than their reference diversity (note different scales). For venues like CSCW or UIST, there is minimal difference between reference and citation diversity; yet the difference is pronounced for venues like the HCI conference. ICCHP has relatively lower reference and citation diversity, suggesting that it is a more niche conference in general. ASSETS and CHI have fairly standard levels of reference diversity and enjoy comparable or higher levels of citation diversity compared to other HCI venues.

\section{Discussion}
\label{discussion}

Citation diversity and measures of interdisciplinarity allow us to comment on the strength of relationships between fields---how often one field cites another or builds upon their work. In this work, we focus on the citational diversity of accessibility research, a subfield of HCI that is cross-disciplinary by nature, as it draws influence from not only the broader HCI community, but also from innovations in rehabilitation medicine, gerontology, psychology, education, and more. For the most part, the venue in which an accessibility paper is published, CHI versus ASSETS, does not affect major differences in a paper's eventual citational impact; the median citation count is similar between the two conferences. Reference and citation patterns between accessibility papers published in the two venues are also similar, perhaps due to an overlap in the authorial community. When applying LCDI over all publications (not just those on accessibility) in ASSETS and CHI, we observe that the relative diversity of ASSETS references and citations are lower than those of CHI. This is unsurprising, since ASSETS is focused on the sub-discipline of accessibility, while CHI represents the broader HCI community. One could conjecture the benefits of both venues: ASSETS focuses on accessibility and papers published there reach a targeted community, while CHI is less thematic but grants exposure to a potentially more diverse research audience among its attendees.

The primary limitation of this study stems from imperfect paper and citation metadata. No database of paper metadata is complete, and we attempt to offset the brunt of this issue by sourcing metadata from two databases. We quantify the bias introduced by data missingness through error analysis, the results of which suggest that there should be minimal impacts to our results. Additionally, citations are only one way in which researchers from different disciplines interact, and they do not fully capture interdisciplinary relationships. Explicit collaborations between authors from different departments, schools, and institutions can also be used as a measure of interdisciplinarity, perhaps in future work. 

Another direction for future work is to explore the nature of venue differences and how they impact individual papers. Though we do not analyze authors in this work, the authorial composition of a paper likely contributes to a paper's reference choices and citation outcomes. Future work could also explore whether a paper's reference diversity is correlated with its citation diversity, i.e., whether a paper that positions itself as more interdisciplinary actually contributes to the furthering of knowledge across broader fields of study.

\section{Conclusion}

Periodic top-down examination of a field's relation to other fields can help the community reflect upon the broader impacts of their work. Our analysis of citation diversity for accessibility papers reveals that though these papers are predominantly influenced by other works in accessibility and HCI, they also draw influence from disability studies, psychology, and other fields. Whether an accessibility paper is published in CHI or ASSETS produces little difference in its citational outcome, though the venues as a whole are rather different. ASSETS exemplifies a more targeted venue, focused on research in accessible computing, while CHI, as a general HCI venue, demonstrates higher reference and citational diversity among its publications. There are complementary benefits to publishing in both venues. We also encourage our fellow researchers to continue drawing inspiration broadly, and look to increasing interdisciplinarity as a way of seeking new avenues for innovation in accessibility.

\begin{acks}
This work was funded in part by the National Science Foundation under grants IIS-1818594 and IIS-1652339 as well as the UW Center for Research and Education on Accessible Technology and Experiences (CREATE). 
\end{acks}

\bibliographystyle{ACM-Reference-Format}
\bibliography{a11y_biblio}


\appendix

\setcounter{table}{0}
\renewcommand{\thetable}{A.\arabic{table}}

\section{Error analysis}
\label{app:error_analysis}

Among the references and citations of the core set without venue information, we analyze the 100 most commonly occurring DOIs. These 100 DOIs are associated with approximately 6\% of all references and citations to venue-less documents. An author manually resolved or attempted to resolve these DOIs, and identified the type of document. A detailed breakdown of document types is given in Table \ref{tab:error_analysis}.

Most of these documents (73 of 100) correspond to books, book chapters, reports, theses, or other forms of documents that are not associated with a publication venue. Of the 100, 14 conference papers and 9 journal articles are associated with a venue that is unknown or missing in our two source databases. Missing conference venues include places like ChineseCSCW, ICDVRAT, and the International Symposium on Signed Language Interpretation and Translation Research. Missing journals include publications like \textit{Social Forces} and \textit{The Journal of Aesthetics and Art Criticism}; notably several of the journals are only available in JSTOR, which may explain the dearth of publicly available metadata. Because all venues seen in this sample are not members of our set of comparative venues, or among any of the top venues in the references or citations of our core or comparative datasets, we believe the degree of missingness over the entire dataset should not dramatically impact our results or interpretation.  

\begin{table}[tbhp!]
    \centering
    \small
    \begin{tabular}{llp{30mm}}
        \toprule
        Type & Uniq. Number \\
        \midrule
        Book chapter & 33 \\
        Book & 21 \\
        Thesis/dissertation & 16 \\
        Conference paper & 14 \\
        Journal article & 9 \\
        Broken DOI & 2 \\
        Other & 2 \\
        Report & 1 \\
        Commentary & 1 \\
        Proceedings & 1 \\
        \bottomrule
    \end{tabular}
    \caption{Error analysis for referenced and citing items without venue information. Of the top 100 most frequently occurring DOI among these items, most are books, book chapters, and theses/dissertations, along with 16.4\% journal and conference articles, and other miscellaneous materials.}
    \label{tab:error_analysis}
\end{table}

\section{CHI and ASSETS references and citations}
\label{app:contingency_tables}

Table~\ref{tab:contingency_tables} provides the contingency tables for the references and citations to and from the CHI and ASSETS subsets.

\begin{table}[h!]
    \small
    \begin{subtable}{0.45\linewidth}
        \centering
        \begin{tabular}{r|cc|l}
            \toprule
            Venue & ASSETS & CHI & Total \\
            \midrule
            CHI & 1238 & 1348 & 2586 \\
            ASSETS & 1180 & 673 & 1853 \\
            \midrule
            Total & 2418 & 2021 & 4439 \\
            \bottomrule
        \end{tabular}
        \caption{Outbound references ($\chi^2 = 108.1$, $p < 0.001$)}
        \label{tab:contingency_refs}
    \end{subtable} 
    \hspace{4mm}
    \begin{subtable}{0.45\linewidth}
        \centering
        \begin{tabular}{r|cc|l}
            \toprule
            Venue & ASSETS & CHI & Total \\
            \midrule
            CHI & 1344 & 1456 & 2800 \\
            ASSETS & 1598 & 799 & 2397 \\
            \midrule
            Total & 2942 & 2255 & 5197 \\
            \bottomrule
        \end{tabular}
        \caption{Inbound citations ($\chi^2 = 182.4$, $p < 0.001$)}
        \label{tab:contingency_cits}
     \end{subtable}
     \caption{Contingency tables for the references and citations of the CHI and ASSETS subsets to the ASSETS and CHI venues. Both references and citations are more likely to extend to and originate from, respectively, papers in the same venues. The chi-square test is significant for both references and citations.}
     \label{tab:contingency_tables}
\end{table}

\end{document}